\documentstyle[twoside,fleqn,espcrc2]{article}
\def\be{\begin{equation}}
\def\ee{\end{equation}}
\def\ba{ \begin{eqnarray}}
\def\ea{ \end{eqnarray}}
\def\nn{\nonumber}

\begin{document}
\title{A\ Self Consistent Study of the Phase Transition in the Scalar
 Electroweak
Theory at Finite Temperature}
\author{U. Kerres, G. Mack \\
II. Institut f\"ur Theoretische Physik
                          der Universit\"at, D22761 Hamburg,
Luruper Chaussee 149\\[1mm]
G. Palma \\
Dpto. de Fisica, Universidad de Santiago de Chile,
Casilla 307, Correo 2, Santiago, Chile.}
\begin{abstract}
We propose the study of the phase transition in the scalar electroweak
theory at finite temperature by a two - step method. It combines i)
dimensional reduction to a 3-dimensional {\it lattice\/} theory
via perturbative
     blockspin transformation, and ii) either further real space
renormalization group transformations, or solution of gap equations,
for the 3d lattice theory. A gap equation can be obtained
by using the Peierls inequality to find the best
quadratic approximation to the 3d action.
This method avoids the lack of self consistency of the usual treatments
which do not separate infrared and UV-problems by introduction of a
lattice cutoff. The effective 3d lattice action could also be used in
computer simulations.
\end{abstract}
\maketitle
\section{Scalar Theory}
To introduce our method \cite{uke}
we consider first the scalar $\lambda
\Phi ^4$-theory on the continuum $\Re ^4$ of points $z$. We define the
blockspin, the effective action,
the fluctuation field propagator and the
interpolation operator. To proceed from temperature $T=0$ to $T>0$,
we use that the finite temperature
propagator $v_{T\mbox{ }}$can be obtained from the zero-temperature
propagator $v_0$ by periodizing in time:
$
v_T(
{\bf r},t) = \sum_{n\in Z}v_0({\bf r},t+n\beta ).
$
%
This reproduces the well known finite temperature propagator
\cite{kap} (sum over Matsubara frequencies $\omega_n=2\pi n$).
We generalize the procedure for the other above mentioned entities.

The continuum is divided into blocks $x$ with extensions $L_s$ in the space
and $L_{t\mbox{ }}$ in the time direction  respectively.
The block centers form a lattice $\Lambda$.              Following Gawedzki
and Kupiainien \cite{gawed} we associate a block spin $\Phi (x)$ to a
scalar field $\phi (z)$ as its block average
$$
\Phi (x)=\,C\,\,\phi (x)=av_{z\in x}\,\,\phi (z) \ .
$$
C is called the averaging operator.

Given the continuum action $S\left[ \phi \right] $, an effective
lattice action $S_{ eff }\left[ \Phi \right] $ is defined in
 Wilson's way
$$
\exp (-S_{eff}\left[ \Phi \right] )=\int {\cal D}\phi \,\delta
(C\phi
-\Phi )\,\exp (-S\left[ \phi \right] )
$$
where in place of the $\delta $-function, one may use a Gaussian
following \cite{hasenfratz}.

Given a block spin $\Phi $,
one determines a background field $\varphi $
which minimizes the free action
$-\frac 12 (\varphi,\Delta \varphi )$
  subject to the constraint $C\varphi = \Phi_{}$. This determines the
interpolation kernel ${\cal A}_T$,
$$
\begin{array}{c}
\varphi (z)=\int_{z\,\in \,\Lambda }\,
{\cal A}_T(z,x)\,\Phi (x),\,\,\mbox{ with }C{\cal A}_T=1 \\
\end{array}
$$
Following \cite{gawed} one splits the field $\phi $ into the  low
frequency part $\varphi $ which
                   is determined by the block spin $\Phi $, and a high
frequency or fluctuation field $\zeta $ which has vanishing block
average, $C\zeta=0$,
and propagator $\Gamma_T = Pv_TP^{\ast}$ with high frequency projector
                                              $P=1-{\cal A}_TC $.
The block spin $\Phi $ has free propagator $u_T=Cv_TC^{\ast}$.
The effective action and its perturbation expansion to all orders
become
\ba
S_{eff}\left[ \Phi \right] &=&\frac 12\left( \Phi ,u_T^{-1}\Phi \right)
+V_{eff}^{}\left[ \Phi \right] +const.\hfill  \label{FT}  \\
 V_{eff}\left[ \Phi \right]&=&-\ln \left( \int d\mu_{\Gamma_T}
      (\zeta )
        \exp \left( -V_\Phi (\zeta )\right) \right)\nn \\
&=&\label{PT}
   \Bigl\{  \exp (\frac{1}{2}\frac{\delta}{\delta \zeta}\Gamma_T
                 \frac{\delta}{\delta \zeta})
\exp \left( -V_{\Phi }(\zeta )\right) \Bigr\}_{\zeta =0}
\\
V_{\Phi }(\zeta )&=& V({\cal A}_T\Phi +\zeta ).
\ea
$d\mu_{\Gamma_T}$ is
 the Gaussian measure with covariance $\Gamma_T$.
Thus $V_{eff}$ is the free energy of a field theory with free
propagator $\Gamma_T $ and $\Phi $-dependent coupling constants.

At finite temperature, time is periodic  with period
$\beta$.
The extension $L_t$ of blocks in time direction must be chosen
commensurate
with $\beta $. $S_{eff}$ depends on $T$, but only weakly so as long as
$\beta>>L_t$.

A great simplification results if we choose $L_t=\beta $, so that only
one block fits in time direction and
  the lattice $\Lambda $ becomes 3-dimensional.

   The   periodic boundary conditions in time direction  are
now inherited by the blocks. As a consequence one finds
          that $u_T=\beta^{-1}u_{FT}$ with a $T$-{\it
independent\/} 3d lattice propagator $u_{FT}$,
                                              so that the kinetic
term in (\ref{FT}) has a factor $\beta$, while $V_{eff}$ has only a
weak $T$-dependence. $\Gamma_T(z,z^{\prime })$ has exponential decay
with decay length of order $L_s$ if $L_s\geq O(L_t)$. Therefore the
effective action is local modulo exponential tails. It can be expanded,
\ba
V_{eff}\left( \Phi \right) =\int_x\left\{ \frac 12m^2\Phi ^2+
 \frac{\lambda_r}{4!}\Phi^4+\frac{\lambda_6}{6!}\Phi^6\right\}
+\hfill \cr
+\frac \beta 2\int_x\int_y\Phi \left( x\right) \,u_{FT}^{-1}(x,y)\Phi
(y)\left[ 2\delta_2+\gamma \Phi (x)^2\right] +... \nn
\ea
where all coefficients are (weakly) T-dependent and finite.
\subsection*{Treatment of the effective theory}
The effective lattice theory can be analysed in different ways.

1) Further block spin transformations can be performed to increase
$L_s$ and ultimately determine the constraint effective potential
\cite{fukuda} which determines the probability distribution of the
magnetization.

2) Solve a gap equation \cite{tak,buchmuller}.
 Following Feynman and Bogoliubov, one may
seek an optimal quadratic approximation $S_0=\frac 12 (\Phi , J \Phi)$
to $S_{eff}$ around which to expand. The optimal choice is that which
maximizes the right hand side of the Peierls inequality
\cite{Ruehl},
$$
\ln Z \geq   \ln Z_0-\left\langle S-S_0\right\rangle _0 .
$$
The optimal $J$ is determined by the extremality condition
$$
\left\langle \frac{\delta ^2S}{\delta \Phi (x)\delta \Phi (y)}\right\rangle
_0=J(x,y) \ .
$$
This
procedure leads to gap equations whose perturbative solution is the
sum of superdaisy diagrams. In principle one can compute corrections,
treating $S-S_0$ as a perturbation.

\subsection*{Comparison with other approaches}
The preceding formulae are also valid when one averages only over time
to obtain a 3-dimensional {\it continuum\/} theory \cite{kaj}.
                                                    In this case
${\cal A}_T= \beta^{-1}{\bf 1}$ while the fluctuation propagator
is translation invariant and given by
$$ \Gamma_T({\bf r},t)= \beta^{-1} \sum_{n\not=0}
    \int \! \frac{d^3{\bf p}}{(2\pi)^3}
e^{-i{\bf p r}-i\omega_nt}[\omega_n^2 +{\bf p}^2]^{-1}  \nn
$$
 The closest singularity of the integrand is at $|{\bf p}|=
\pm 2\pi i \beta^{-1}$. Therefore $\Gamma_T$ decays exponentially
with ${\bf r}$ with decay length $\beta/2\pi$. According to
the perturbation expansion
eq.(\ref{PT}) this leads to nonlocalities in $S_{eff}$ in higher orders.
This is a disaster. In contrast with the lattice case, derivative
expansions of the nonlocal terms are not possible here because they
lead to UV-divergences.

An inconsistency is also encountered in the
standard 4-dimensional approach based on a gap equation
for the continuum theory whose formal
solution is the sum over super daisy diagrams.
It is supposed to determine $m^2(T)$.          If static and nonstatic
contributions are treated on the same footing, one finds
a {\it temperature dependent}
logarithmically divergent contribution to the self energy $\Pi $:
$$
\Pi _{vac}=\frac 1{16\pi ^2}\left[ \Lambda ^2-m^2(T)\ln (\frac{\Lambda ^2}{%
m^2(T)})\right] +O(\frac 1{\Lambda ^2})\mbox{ }
$$
To cancel the logarithmic UV-divergence a temperature dependent
counterterm is needed. This is not acceptable.
Our method avoids this by separating UV and IR problems.
\section{Scalar Electrodynamics}
In this section we outline the extension of the previous sections to
scalar electrodynamics.
%
%
We recall first the Balaban-Jaffe block spin transformation for the
abelian gauge field \cite{balimja}

Given a vector potential $a(z)=a_\mu (z)dz_\mu $ on the continuum, the block
spin $A$ is defined through
$$
A\left[ b\right] =A_\mu (x)=av\,_{z\,\in \,x\,}\int_{C_{z,\mu }}d\omega _\nu
\,\,a_\nu (\omega )\mbox{ }
$$
where $b$ is the link emanating from $x$ in $\mu $-direction, $z$ a point $%
z\in \,x$ and $C_{z,\mu }$ the straight path of length one block lattice
spacing in $\mu $-direction starting from $z.$

If $\mu =4$ , $C_{z,\mu }$ connects $z$ with $z+L_te_4,$ and if $\mu \neq 4$
it connects $z$ with $z+L_se_\mu $ ($e_\mu $ = unit vector in $\mu $%
-direction)

The blocking procedure has to be
                       covariant under gauge transformations.
%
%
Given a Higgs field $\phi (z),$ we wish to define a covariant block Higgs
field $\Phi (x)$ . In order to maintain gauge covariance, one must use
an  averaging kernel which
   depends  on the gauge field $a.$ Our proposal \cite{mack} is to
define the averaging operator for the Higgs field $C^H(a)$
as lowest eigenmode of the
           covariant Laplacian $\Delta_a$ with Neumann boundary
conditions
on the block $x:$
$$
-\Delta _a^{N,x}\,C^{H{\dagger }}(a\mid z,x)=\epsilon _0(a\mid
x)\,C^{H{\dagger }}(a\mid z,x)
$$
while $C^H(a\mid z,x)=0$ for
 $z\notin x$, and $C^HC^{H{\dagger}}=1$.
 $\epsilon _0(a\mid x)$ is the lowest eigenvalue of $-\Delta_a^{N,x}$
  and $C^{H{\dagger }}(a\mid z,x)$ is the adjoint kernel.
$C^H$ admits a perturbation expansion \cite{palma}.

The  choice $L_t=\beta$ leads again to a 3-di\-men\-sional lattice.
Periodic boundary conditions on blocks in time
direction  are again imposed. The solution of the eigenvalue
equation for the averaging kernel $C^H$
                                  for the Higgs field will depend on
$T$.
%
%
The effective 3-dimensional lattice action for scalar
electrodynamics is
\ba
S_{eff}(\Phi ,A)=-\ln \int {\cal D}a\int {\cal D}\phi \,\delta
_{Ax}(a)\delta (Ca-A)  \hfill \cr  \hfill
\delta (C^H(a)\phi -\Phi )\,e^{-S_M(a)-\frac 12(\phi ;\Delta _a\phi )-V(\phi
)} \nn
\ea
where $V(\phi )=\int dz\left[ \frac 12m_0^2(\phi \phi ^{\dagger })+\frac
\lambda {4!}(\phi \phi ^{\dagger })^2\right] $, $S_M(a)$ is the Maxwell
action, and $\delta _{Ax}(a)$ is the block radial gauge fixing term, which
fixes the gauge only locally within each block \cite{balimja}.

After transformation to a block Landau gauge, $S_{eff} $ can be computed
by perturbation theory as for the scalar case. The Faddeev Popov
                                                   transformation
to the block Landau gauge as well as the interpolation kernel
${\cal A}_T$ and the fluctuation propagator $\Gamma_T$
 were discussed in detail by Balaban and Jaffe \cite{balimja}.
The only new aspect in our situation is the existence of periodic
b.c. on blocks in time direction.

This work was supported in part by DICYT \#049331PA, FONDECYT \#1930067 and
DFG.

GP wants to thank II. Institut f\"ur theoretische Physik in
Hamburg
for the kind hospitality during the end phase of this work.

\end{document}